\documentclass{llncs}
\usepackage[utf8]{inputenc}
\usepackage[margin=1in]{geometry}  
\usepackage{graphicx}              
\usepackage{amsmath}               
\usepackage{amsfonts}              
\title{PH = PSPACE}
\titlerunning{PH = PSPACE}  
\author{Valerii Sopin\orcidID{0000-0001-9953-6072}}
\authorrunning{V. Sopin} 
\tocauthor{Valerii Sopin}
\institute{
\email{VvS@myself.com}, \texttt{ https://www.researchgate.net/profile/Valerii-Sopin}}

\date{2022}

\begin{document}

\maketitle              

\begin{abstract}
We show that $\mathbb{PSPACE}$ is equal to 4th level in the polynomial hierarchy. 

\textbf{MSC classes}:	03G05, 03B70, 68Q12, 68Q15, 68Q25
\end{abstract}

\section{Introduction}
In computational complexity theory, $\mathbb{NP}$ is one of the most
fundamental complexity classes. The complexity class $\mathbb{NP}$
is associated with computational problems having solutions that,
once given, can be efficiently tested for validity. It is customary
to define $\mathbb{NP}$ as the class of languages which can be
recognized by a non-deterministic polynomial-time machine.

A decision problem is a member of co-$\mathbb{NP}$ if and only if
its complement (the complement of a decision problem is the decision
problem resulting from reversing the "yes" and "no" answers) is in
the complexity class $\mathbb{NP}$. In simple terms,
co-$\mathbb{NP}$ is the class of problems for which efficiently
verifiable proofs of "no" instances, sometimes called
counterexamples, exist. Equivalently, co-$\mathbb{NP}$ is the set of
decision problems where the "no" instances can be accepted in
polynomial time by a non-deterministic Turing machine.

On the other hand, $\mathbb{PSPACE}$ is the set of
all decision problems that can be solved by a Turing machine using a
polynomial amount of space.

An oracle machine is an abstract
machine used to study decision problems. It can be visualized as a
Turing machine with a black box, called an oracle, which is able to
solve certain decision problems in a single operation. We use notation
$\mathbb{L}^{\mathbb{O}}$, where $\mathbb{O}$ is the
oracle.

On the contemporary state-of-the-art, the interested reader is
referred to [1] and references therein.

Our result resolves some of unsolved problems in Computer Science. 

The essential idea of the proof is to show that for any (fully) quantified Boolean formula $\phi$ we can obtain a formula $\phi'$ which is in the fourth level of the polynomial hierarchy, no more than polynomial in the size of a given $\phi$, such that the truth of $\phi$ can be determined from the truth of $\phi'$. The idea is to skolemize, and then use additional formulas from the second level of the polynomial hierarchy inside the skolemized prefix to enforce that the skolem variables indeed depend only on the universally quantified variables they are supposed to. However, some dependence is lost when the quantification is reversed. It is called "XOR issue" in the paper because the functional dependence can be expressed by means of an XOR formula. Thus, it is needed to locate these XORs. The last can be done locally, when all arguments are specified (keep in mind the algebraic normal form (ANF)), i.e. as a polynomial subroutine.

The paper is organized as follows. Chapters 2-4 refresh basic definitions. Chapter 5 contains the proof. 

\section{Quantified Boolean formula}
The Boolean Satisfiability Problem (abbreviated as SAT) is the
problem of determining if there exists an interpretation that
satisfies a given Boolean formula. In other words, it asks whether
the variables of a given Boolean formula can be consistently
replaced by the values true or false in such a way that the formula
evaluates to true. SAT was the first known $\mathbb{NP}$-complete problem, as proved by Stephen Cook [2] and independently by Leonid Levin [3].

One simple example of a co-$\mathbb{NP}$-complete problem is
tautology, the problem of determining whether a given Boolean
formula is a tautology; that is, whether every possible assignment
of true/false values to variables yields a true statement.

For a Boolean formula $\phi(x_1,\dots,x_n)$, we can think of its
satisfiability as determining the true of the statement
$$\exists x_1\in \{0,1\}\; \exists x_2\in \{0,1\}\; \dots\; \exists x_n\in
\{0,1\}\;\; \phi(x_1,\dots,x_n).$$

The SAT problem becomes more difficult if both "for all" $(\forall)$
and "there exists" $(\exists)$ quantifiers are allowed. It is known
as the quantified Boolean formula problem or QSAT. QSAT is the
canonical complete problem for $\mathbb{PSPACE}$ [1].

\section{The Polynomial Hierarchy}
We have seen the classes $\mathbb{NP}$ and co-$\mathbb{NP}$, which
are defined as follows [1]:

$L \in \mathbb{NP}$ if there is a deterministic Turing machine $M$
running in time polynomial in its first input, such that $x \in L
\Leftrightarrow \exists w\; M(x;w) = 1,$ $w$ has length
polynomial in $x$.

$L \in $ co-$\mathbb{NP}$ if there is a deterministic Turing machine
$M$ running in time polynomial in its first input, such that $x \in
L \Leftrightarrow \forall w\; M(x;w) = 1,$ $w$ has length
polynomial in $x$.

It is natural to generalize the above [1][4].

Let $i$ be a positive integer. $L \in \Sigma_i$ if there is a
deterministic Turing machine $M$ running in time polynomial in its
first input, such that $$x \in L \Leftrightarrow \underbrace{\exists
w_1 \forall w_2 \dots Q_i w_i}_{i \text{ times }}\; M(x; w_1; \dots;
w_i) = 1,$$ where $Q_i =\forall$ if $i$ is even, and $Q_i = \exists$
if $i$ is odd.

Let $i$ be a positive integer. $L \in \Pi_i$ if there is a
 deterministic Turing machine $M$ running in time polynomial in its
first input, such that $$x \in L \Leftrightarrow \underbrace{\forall
w_1 \exists w_2 \dots Q_i w_i}_{i \text{ times }}\; M(x; w_1; \dots;
w_i) = 1,$$ where $Q_i =\forall$ if $i$ is odd, and $Q_i = \exists$
if $i$ is even.

As in the cases of $\mathbb{NP}$, co-$\mathbb{NP}$, we require that  $w_i$ each have length polynomial in $x$.

The polynomial hierarchy $\mathbb{PH}$ consists of all those
languages of the form defined above. Note also the similarity to QSAT. The crucial difference is that QSAT allows an unbounded number of alternating quantifiers, whereas $\Sigma_i$, $\Pi_i$ each allow (at most) $i$ quantifiers. From here, $\mathbb{PH} \subseteq \mathbb{PSPACE}$.

\section{Alternating Turing machine}
An alternating Turing machine (ATM) is a non-deterministic Turing
machine (NTM) with a rule for accepting computations that
generalizes the rules used in the definition of the complexity
classes $\mathbb{NP}$ and co-$\mathbb{NP}$. The concept of an ATM
was set forth by Ashok Chandra, Larry Stockmeyer and Dexter Kozen [5].

The definition of $\mathbb{NP}$ uses the existential mode of
computation: if any choice leads to an accepting state, then the
whole computation accepts. The definition of  co-$\mathbb{NP}$ uses
the universal mode of computation: only if all choices lead to an
accepting state, then the whole computation accepts. An alternating
Turing machine (or to be more precise, the definition of acceptance
for such a machine) alternates between these modes.

An alternating Turing machine with $k$ alternations is an
alternating Turing machine which switches from an existential to a
universal state or vice versa no more than $k-1$ times. The
complexity class $\mathbb{PH}$ is a special case of hierarchy of
bounded alternating Turing machine [5].

$\mathbb{AP} = \mathbb{PSPACE}$, where $\mathbb{AP}$ is the class of
problems alternating machines can solve in polynomial time [5].

\section{Main result}

Next theorem shows that QBF is indeed a \textbf{generalisation} of the Boolean Satisfiability Problem, where determining of interpretation that satisfies a given Boolean formula is replaced by existence of Boolean functions that makes a given QBF to be  tautology. Such functions are called the \textbf{Skolem functions}.

\begin{theorem}The quantified Boolean formula $$\Omega_1 x_1\in
\{0,1\}\; \Omega_2 x_2\in \{0,1\}\; \dots \;\Omega_n x_n\in \{0,1\}
\;\;\phi(x_1,\dots,x_n),$$ where $\phi(x_1,\dots,x_n)$ is a  Boolean formula, $\Omega_{s}$, $s = i_1, \dots, i_j,$ is the quantifier
$\exists$ and $\Omega_{t}$, $t \neq i_1, \dots, i_j,$ is the quantifier
$\forall$, $j$ is the number of variables with the quantifier
$\exists$, is a true quantified Boolean formula if and only if there are Boolean functions $y_q$, where $y_q$ depends only on variables with the quantifier $\forall$ and indexes less $i_q$, $ q = 1, \dots, j,$ that after substituting $x_{i_q} := y_q$  the given quantified Boolean formula becomes tautology.\end{theorem}

\begin{proof}  It follows from a simple recursive
algorithm for determining whether a QBF is true. We take off the first quantifier and check both possible values for the first variable:
$$A = \Omega_2 x_2\in \{0,1\}\; \dots \;\Omega_n x_n\in \{0,1\}
\;\;\phi(0,\dots,x_n),$$
$$B = \Omega_2 x_2\in \{0,1\}\; \dots \;\Omega_n x_n\in \{0,1\}
\;\;\phi(1,\dots,x_n).$$

If $\Omega_1 = \exists$, then return $A$ disjunction $B$ (that's it, $A$ or $B$ is true; to avoid unambiguous, if $A$ and
$B$ is true, take $A$ for determining the function, so the value depends only on values of previous variables). If $\Omega_1 = \forall$, then return $A$ conjunction $B$ ($A$ and $B$ is true).

Notice that a Boolean function determines the truth table (one-to-one correspondence).\end{proof} 

\begin{example} Let only the quantifier for $x_k$, $k \geq 1$, be existential, then $y_1$ is some function of variables $x_1, \dots, x_{k-1}$, as QBF means in that case that for any possible values of $x_1, \dots, x_{k-1}$ there exists value of $x_k$ that for all possible values of $x_{i>k}$ the given formula is true. It is indeed the truth table, where values of $x_1, \dots, x_{k-1}$ determine the value $x_k$.
\end{example}

\begin{example} $\forall x_1 \exists z_1 \forall x_2 \exists z_2 \forall x_3 \exists z_3\; \phi(x_1, z_1, x_2, z_2, x_3, z_3)$ is a true QBF if and only if there exist such Boolean functions $y_1: \lbrace 0, 1\rbrace \rightarrow \lbrace 0, 1\rbrace$, $y_2: \lbrace 0, 1\rbrace^2 \rightarrow \lbrace 0, 1\rbrace$, $y_3: \lbrace 0, 1\rbrace^3 \rightarrow \lbrace 0, 1\rbrace$ that $$\phi(x_1, y_1(x_1), x_2, y_2(x_1, x_2), x_3, y_3(x_1, x_2, x_3)) \text{ is tautology.}$$ \end{example}

\begin{theorem} $$\prod\nolimits_4 = (\text{co-}\mathbb{NP})^{\mathbb{NP}^{(\text{co-}\mathbb{NP})^{\mathbb{NP}}}} =  \mathbb{PSPACE}$$\end{theorem}

\begin{proof}  From [1][6] we know that without loss of generality we can assume a  quantified Boolean formula to be in form (prenex normal form), where existential and universal quantifiers alternate.  We assume it, for simplicity.

We \textbf{wish} that a quantified Boolean formula $$\forall x_1\in
\{0,1\}\; \exists y_1 \in \{0,1\}\; \forall x_2\in
\{0,1\}\; \exists y_2 \in \{0,1\}\; \dots \;\forall x_n \in
\{0,1\}\; \exists y_{n}\in \{0,1\}$$ $$
\;\;\phi(x_1, y_1, \dots, x_n, y_{n})$$ would be equivalent (equisatisfiable, more correctly) to\\
$\forall (x_1, x_2, \dots, x_{n})$ $\exists( y_1, \dots, y_n) \lbrace$
$$\phi(x_1, y_1, x_2, y_2, \dots, x_{n}, y_n) \; \wedge \; $$
$$\wedge\;  \forall  (\hat{x}_{n})\;  \exists (z_n)\;\;  \phi(x_1, y_1, x_2, y_2, \dots, x_{n-1}, y_{n-1}, \hat{x}_{n}, z_n)\;  \wedge \;$$
$$\wedge\; \forall  (\hat{x}_{n-1}, \hat{x}_{n}) \; \exists (z_{n-1}, z_n) \;\; \phi(x_1, y_1, x_2, y_2, \dots, x_{n-2}, y_{n-2},\hat{x}_{n-1}, z_{n-1}, \hat{x}_{n}, z_n) \; \wedge \; \dots$$
$$\dots \;\wedge\; \forall  (\hat{x}_{2},\dots, \hat{x}_{n}) \;  \exists (z_{2},\dots,  z_n) \;\; \phi(x_{1}, y_1,\hat{x}_{2}, z_2, \dots, \hat{x}_{n-2}, z_{n-2}, \hat{x}_{n-1}, z_{n-1}, \hat{x}_{n}, z_n)$$
$$\rbrace\;\;\;\;\;\;\;\;\;\;\;\;\;\;\;\;\;\;\;\;\;\;\;\;\;\;\;\;\;\;\;\;\;\;\;\;\;\;\;\;\;$$

Namely, iterations of $\forall x \exists y$ reduce to  conjunctions of separated $\forall \hat{x} \exists z$, as in the beginning we fix values of $\{y_q, \; q=1, \dots, n\}$ and conjunctions jointly check that for predetermined $\{y_l,\; l< q\}$ suitable continuation $\{y_l,\; l \geq q\}$ can be found. In each conjunction we consider $\{y_l, l < q\}$ as functions dependent on all $\{x_i, \; i<q\}$  and $\{z_l, l \geq q\}$  as functions dependent on every $\{x_i, \; i=1, \dots, n\}$ (if $\forall x_1\;  F(x_1, 0) = F(x_1, 1),$ then variable $x_2$ is dummy variable for Boolean formula $F(x_1, x_2)$). From here, if it is a true quantified Boolean formula, the above confirms it. However, another implication is not always true. Let's exam when two parts are different, allowing $\phi$ to have also odd number of variables with preserving alternations for quantifiers for foregoing induction.

$\textbf{m $=$ 1:}$ for a Boolean formula of one variable  the equivalence  obviously holds.

$\textbf{m $=$ 2:}$ inconsistency can possibly happen only with $\exists y \; \forall x \;  \phi(y, x)$; we have 16 different Boolean formulas of two variables and the equivalence is violated only for $\textsl{XOR}: (y \oplus x), \neg (y \oplus x).$

\begin{example}  $\exists y \; \forall x \;\; x \cdot y$ is FALSE as well as $\forall x \;  \exists y \;\;  x \cdot y$\end{example}

\begin{example} 
        $\exists y \; \forall x \;\; x+ y$ is TRUE as well as $\forall x \;  \exists y \;\;  x + y$\end{example}

\begin{example} 
       $\exists y \; \forall x \;\; x \oplus y$ is FALSE, but $\forall x \;  \exists y \;\;  x \oplus y$ is TRUE\end{example}

$\textbf{m $\geq$ 3:}$ taking off the first quantifier and checking both possible values for the first variable in way we did in Theorem 1, we come to the $\textbf{m - 1}$ case. Indeed, for example, considering $\textbf{m $=$ 3}$, we have 

$$\forall z \; \exists y \; \forall x \;\; \phi(z, y, x) \equiv \exists y \; \forall x \;\; \phi(0, y, x) \; AND \; \exists y \; \forall x \;\; \phi(1, y, x),$$
$$\;\;\; \exists t \; \forall x \;  \exists y \;\; \phi(t, x, y) \equiv \forall x \; \exists y \;\; \phi(0, x, y) \; OR \;  \forall x \; \exists y \;\; \phi(1, x, y),$$
where the second expression can be viewed as negation of the first expression. Consequently, it is enough to inspect only the first expression due to double negation.

If $\exists y \; \forall x \;\; \phi(0, y, x) \equiv \forall x \; \exists y \;\; \phi(0, y, x)$ and $\exists y \; \forall x \;\; \phi(1, y, x) \equiv \forall x \; \exists y \;\; \phi(1, y, x)$, then  $\forall z \; \exists y \; \forall x \;\; \phi(z, y, x)  \equiv  \forall z \; \forall x \; \exists y \;\; \phi(z, y, x) \equiv  \forall z \; \exists \xi \; \forall x \; \exists y \;\; \phi(z, y, x)$. Otherwise, the equivalence is false due to $\textsl{XOR}$ issue from $\textbf{m $=$ 2}$. Then $\exists y \; \forall x \;\; \phi(0, y, x)$ or $\exists y \; \forall x \;\; \phi(1, y, x)$ is false. Therefore, $\forall z \; \exists y \; \forall x \;\; \phi(z, y, x)$ is false.

Thus, using mathematical induction we have shown that XOR issue from $\textbf{m $=$ 2}$ appears whenever the equivalence we want doesn't work and the emergence means that the real value is false, but the displayed formula says that it is true (there is no need to locate all chains with XORs: any chain includes a XOR of only two variables).  The algebraic normal form (ANF, Zhegalkin normal form) is used here, i.e. the fact that any Boolean formula can be rewritten using only conjunctions and XORs. 

Indeed, as we have $2n$ variables, XOR issue from $\textbf{m $=$ 2}$ appears not only lonely, it can be some polynomial from ANF representation, i.e., $$(x_1\;  AND \;  \dots \; AND \; x_s \; AND \; x)\;  XOR \;  (y_1 \;  AND \; \dots\;  AND \; y_t \; AND \; y) \;  XOR \; \dots,$$ which is needed to be detected. However, despite the exponential number of such polynomials, the $AND$ allows to examine them as a polynomial subroutine.

So, for

$\;\;\;\;\;\;\;\;\;\;\;\;\; \forall (x_1, x_2, \dots, x_{n})$ $\exists( y_1, \dots, y_n)$ $\forall  (\hat{x}_{i},\dots, \hat{x}_{n}) \;  \exists (z_{i},\dots,  z_n) \;\;$ $$\;\;\;\;\;\;\;\;\;\;\;\;\;\;\;\;\;\;\;\;\;\;\;\;\;\;\;\;\;\;\;\;\;\;\;\;\;\;\;\;\;\;\;\;\;\;\;\;\;\;\;\;\;\;\;\;\;\;\;\;\;\;\;\;\;\;\;\;\;\;\;\;\;\;\;\;\;\;\;\;\;\;\;\;  \phi(x_1, y_1, x_{2}, y_2, \dots, \hat{x}_{i}, z_{i}, \dots, \hat{x}_{n-1}, z_{n-1}, \hat{x}_{n}, z_n)$$
we additionally need to verify that for specific $x_1, \dots, x_{i-1}$ and $\hat{x}_{i},\dots, \hat{x}_{n}$ and found $y_1, \dots, y_{i-1}$ and $z_{i},\dots,  z_n$ the given formula $\phi$ with the above fixed arguments except any variable with universal quantifier and any variable with existential quantifier is not equivalent to $\exists y \; \forall x \; (x \oplus y)$ or $\exists y \; \forall x \; \neg (x \oplus y)$ (there are $n^2$  formulas to examine in total). 

The idea is the following:  if something1 XOR something2 XOR $\dots$ causes unequisatisfiability, then for certain values of variables  the expression something1 XOR something2 XOR $\dots$ must become $\exists y \; \forall x \; (x \oplus y)$ or $\exists y \; \forall x \; \neg (x \oplus y)$. And, otherwise, if something1 XOR something2 XOR $\dots$ does not cause unequisatisfiability, then there are not  values of variables such that the expression something1 XOR something2 XOR $\dots$ becomes to be in the form $\exists y \; \forall x \; (x \oplus y)$ or $\exists y \; \forall x \; \neg (x \oplus y)$. This finds some analogy with Full (Perfect) Disjunctive Normal Form.

To conclude, definition of alternating Turing machine shows that $(\text{co-}\mathbb{NP})^{\mathbb{NP}^{(\text{co-}\mathbb{NP})^{\mathbb{NP}}}}$ is enough and this way we solve complete problem for $\mathbb{PSPACE}$.\end{proof} 

\begin{remark} $\mathbb{PSPACE}$ $=$ $\mathbb{P}$? $\mathbb{PSPACE}$ $=$ $\mathbb{NP}$? $\mathbb{PSPACE}$ $=$ $\mathbb{P}^{\mathbb{NP}}$? $\mathbb{PSPACE}$ $=$ $\mathbb{NP}^{\mathbb{NP}}$? $\mathbb{PSPACE}$ $=$ $\mathbb{NP}^{\mathbb{NP}^{\mathbb{NP}}}$?  \end{remark}

\begin{remark} $\textit{Maximal Satisfying Assignment}_{odd}$, the problem of indicating, that the lexicographical maximum $x_1,  \dots, x_n \in \lbrace 0, 1\rbrace^n$, that satisfies a given Boolean formula, is odd (is $x_1$ odd?), is complete for $\mathbb{P}^{\mathbb{NP}}$.\end{remark}

$\mathbb{BQP}$ (bounded-error quantum polynomial time) is the class of decision problems solvable by a quantum computer in polynomial time, with an error probability of at most $1/3$ for all instances, see [1][6].

\begin{corollary} The polynomial hierarchy collapses and $\mathbb{BQP}$ $\subseteq$ $\mathbb{PH}$.\end{corollary}

\begin{proof}  See Chapter 3 and Theorem 2. It is known that $\mathbb{BQP}$ $\subseteq$ $\mathbb{PSPACE}$.\end{proof} 

\begin{remark} The relationship between $\mathbb{BQP}$ and $\mathbb{PH}$ has been an open problem since the earliest days of quantum computing
$[7]$.\end{remark}

\begin{remark} The polynomial hierarchy is infinite  relative to a random oracle with probability 1 and there exists an oracle separation of $\mathbb{PH}$ and $\mathbb{PSPACE}$, see, for example, $[8]$. However, note that an oracle separation does not necessarily imply the ordinary separation. There is no contradiction.

The proof of Theorem 2 relies strongly on Boolean algebra (the exchange is possible due to finite possibilities for arguments) and that defeats \textbf{the relativization}. Moreover, one useful reformulation is that $\mathbb{PH}= \mathbb{PSPACE}$ if and only if second-order logic over finite structures gains no additional power from the addition of a transitive closure operator. \end{remark}

$\mathbb{BPP}$ (bounded-error probabilistic polynomial time) is the class of decision problems solvable by a probabilistic Turing machine in polynomial time with an error probability bounded away from $1/3$ for all instances, see [1]. If the access to randomness is removed from the definition of $\mathbb{BPP}$, we get the complexity class $\mathbb{P}$.

\begin{corollary} If $\mathbb{P}$ $=$ $\mathbb{NP}$, then $\mathbb{P}$ = $\mathbb{PSPACE}$. If $\mathbb{BPP}$ $=$ $\mathbb{NP}$, then $\mathbb{BPP}$ = $\mathbb{PSPACE}$.\end{corollary}

\begin{proof}  If $\mathbb{P}$ = $\mathbb{NP}$, then $\mathbb{NP}$ = co-$\mathbb{NP}$, since $\mathbb{P}$ = co-$\mathbb{P}$. Moreover, a $\mathbb{P}$ machine with the power to solve $\mathbb{P}$ problems instantly (a $\mathbb{P}$ oracle machine) is not any more powerful than the machine without this extra power. Thus, we obtain that $\mathbb{P}$ = $\mathbb{PH}$.  

$\mathbb{BPP}$ can be treated in the same manner, as it is known that $\mathbb{BPP}$ is closed under complement and  low for itself, meaning that $\mathbb{BPP}^{\mathbb{BPP}} = \mathbb{BPP}$.\end{proof} 

\begin{corollary} If $\mathbb{NP}$ $=$ $\text{co-}\mathbb{NP}$, then $\mathbb{NP}$ = $\mathbb{PSPACE}$.\end{corollary}
\begin{proof}  It is known that if $\mathbb{NP}$ $=$ $\text{co-}\mathbb{NP}$, then $\mathbb{NP}$ = $\mathbb{PH}$.\end{proof} 

$\mathbb{PP}$ is the class of decision
problems solvable by a probabilistic Turing machine in polynomial
time, with an error probability of less than $1/2$ for all
instances, see [1][9]. $\mathbb{PP}$ has natural complete problems, for example, MAJSAT. It is a decision problem, in which one is given a Boolean formula $\phi$. The answer must be "yes" if more than half of all assignments make $\phi$ true and "no" otherwise. 

\begin{corollary} $\mathbb{P}^{\mathbb{PP}}$ $=$ $\mathbb{PSPACE}$.\end{corollary}
\begin{proof} By Toda's theorem $\mathbb{PH} \subseteq \mathbb{P}^{\mathbb{PP}}$  [1][10]. Further, $\mathbb{P}^{\mathbb{PP}} \subseteq \mathbb{P}^{\mathbb{PSPACE}} = \mathbb{P}^{\mathbb{PH}} = \mathbb{PH}$.\end{proof} 

\begin{remark} By adding postselection to $\mathbb{BQP}$ $(\mathbb{BQP}$ $\subseteq$  $\mathbb{PP})$, a larger class is obtained $[11]$. It is known that it is equal to $\mathbb{PP}$ $[11]$. Is it true that $\mathbb{BQP}$ $\neq$ $\mathbb{PP}$, $\mathbb{PP}$ $\neq$ $\mathbb{PSPACE}$?\end{remark}

\begin{corollary} If $\mathbb{NP}$ $\subseteq$ $\mathbb{BQP}$, then $\mathbb{BQP}$  $=$ $\mathbb{PSPACE}$.\end{corollary}
\begin{proof} $\mathbb{BQP}$ is low for itself, which means $\mathbb{BQP}^{\mathbb{BQP}}$ $=$ $\mathbb{BQP}$ [12];  $\mathbb{BQP}$ $\subseteq$ $\mathbb{PSPACE}$.\end{proof}

\begin{remark} Dependency quantified Boolean formulas (DQBFs) are a generalization of ordinary quantified Boolean formulas $[13]$. While the latter is restricted to linear dependencies of existential variables in the quantifier prefix, DQBFs allow arbitrary dependencies, which are explicitly specified in the formula.  This makes decision problem with a DQBF to be $\mathbb{NEXP}$-complete $[14]$. 

Theorem 2 is not applicable to the case of DQBFs directly as the looping is possible (\textbf{the linear order} is used in Theorem 2). Is it within reach to generalise Theorem 2 for it? Notice that $\mathbb{NEXP} \subseteq \mathbb{EXP}^{\mathbb{NP}}$.\end{remark}

\begin{remark} Theorem 2 opens the road for comprehensive pursuing of all exponential complexity classes and their relationships with probabilistic Turing machines and the polynomial hierarchy. The beginning of such kind of research can be found in $[15][16][17][18][19]$.\end{remark}

\begin{corollary} $\mathbb{PSPACE}$ $\neq$ $\mathbb{EXP}$.\end{corollary}
\begin{proof} $\mathbb{EXP}$ $\neq$  $\mathbb{EXP}^{\mathbb{EXP}}$  by the time hierarchy theorems, but $\mathbb{PSPACE}^{\mathbb{PSPACE}} =\mathbb{PH}^{\mathbb{PH}} = \mathbb{PH}$ due to Theorem 2.\end{proof} 

\begin{remark} The complexity of the ATL (Alternating-time Temporal Logic) satisfiability problem was proven to be $\mathbb{EXP}$-complete by van Drimmelen $[20]$ for a fixed number of agents (notice that even with an unbounded supply of agents it is true $[18]$ and that gives another insight about collapsing of the polynomial hierarchy). 

There is fixed point representation of ATL via QBF encoding, i.e., Unbounded Model Checking, see $[21]$. Hence, there exists a translation of ATL formulas into propositional formulas. 

According to Remark 6 and Corollary 6 decision problem with a certain type DQBF is $\mathbb{EXP}$-complete. What could possibly be in-between these extremes? Notice that  if $\mathbb{EXP} \neq \mathbb{NEXP}$, then $\mathbb{P}$ $\neq$ $\mathbb{NP}$.\end{remark}

$$\textbf{Multiset $\lbrace\mathbb{P}$, $\mathbb{NP}$, $\mathbb{NP}^{\mathbb{NP}}$, $\mathbb{NP}^{\mathbb{NP}^{\mathbb{NP}}}$, $\mathbb{NP}^{\mathbb{NP}^{\mathbb{NP}^{\mathbb{NP}}}}\rbrace$ shows that there is always a key.}$$

\section{Acknowledgments}
The author would like to thank Lew Gordeew $[22][23]$,  Zhaohui Wei, Emil Jeřábek, James Cook and the anonymous reviewers for their comments and suggestions.




\newpage
\begin{thebibliography}{1}

\bibitem{1} S. Arora and B. Barak, \textit{Computational Complexity: A Modern Approach}, Cambridge University
Press, 2009.

\bibitem{2} S. Cook, \textit{The Complexity of Theorem-Proving Procedures}, Proceedings of the 3rd Annual ACM Symposium on Theory of Computing, 1971, 151 -158.

\bibitem{3} L. Levin, \textit{Universal search problems}, Problems of Information Transmission, \textbf{9}:3, 1973, 115-116.

\bibitem{4} L. Stockmeyer, \textit{The polynomial-time hierarchy}, Theoretical Computer Science, \textbf{3}, 1977, 1-22.

\bibitem{5} A. Chandra, D. Kozen, L. Stockmeyer, \textit{Universal search problems},  Journal of the ACM, \textbf{28}:1, 1981,  114-133.

\bibitem{6} R. Jain, J. Ji, S. Upadhyay, \textit{QIP = PSPACE},  Proceedings of the 42nd ACM symposium on Theory of computing, 2010, 573--582.

\bibitem{7} S. Aaronson, \textit{BQP and the Polynomial Hierarchy}, Proceedings of the 42nd Annual ACM Symposium on Theory of computing, 2010, 141--150.

\bibitem{8}  R. Raz, A. Tal, \textit{Oracle Separation of BQP and PH}, https://eccc.weizmann.ac.il/report/2018/107/, 2018.

\bibitem{9} J. Gill, \textit{Computational complexity of probabilistic Turing machines}, SIAM Journal on Computing, \textbf{6}:4, 1997, 675--695.

\bibitem{10} S. Toda, \textit{PP is as hard as the polynomial-time hierarchy}, SIAM Journal on Computing, \textbf{20}:5, 1991, 865--877.

\bibitem{11} S. Aaronson, \textit{Quantum computing, postselection, and probabilistic polynomial-time}, Proceedings of the Royal Society A.,\textbf{461}:2063, 2005, 3473--3482.

\bibitem{12} E. Bernstein, U. Vazirani, \textit{Quantum Complexity Theory}, SIAM Journal on Computing, \textbf{26}:5, 1997, 1411--1473.

\bibitem{13} G. Peterson, J. Reif, \textit{Multiple-person alternation},  20th Annual Symposium on Foundations of Computer Science, 1979, 348--363.

\bibitem{14} G. Peterson, J. Reif, S. Azhar, \textit{Lower bounds for multiplayer non-cooperative games of incomplete information},  Computers and Mathematics with Applications, \textbf{41}:7-8, 2011, 957--992.

\bibitem{15} L. Babai, L. Fortnow, N. Nisan and A. Wigderson, \textit{BPP has subexponential time simulations unless EXPTIME has publishable proofs}, Computational Complexity, \textbf{3}, 1993, 307--318.

\bibitem{16} R. Impagliazzo and A. Wigderson, \textit{P = BPP if E requires exponential circuits: Derandomizing the XOR Lemma}, Proceedings of the Twenty-Ninth Annual ACM Symposium on Theory of Computing, 1997, 220--229.

\bibitem{17} S. Mocas, \textit{Separating classes in the exponential-time hierarchy from classes in PH}, Theoretical Computer Science,  \textbf{158}, 1996, 221--231.

\bibitem{18} D. Walther, C. Lutz, F. Wolter, M. Wooldridge, \textit{Atl satisfiability is indeed ExpTime-complete}, Journal of Logic and Computation, \textbf{16}, 2006, 765--787.

\bibitem{19} S. Schewe, \textit{ATL* Satisfiability Is 2EXPTIME-Complete}, International Colloquium on Automata, Languages, and Programming, \textbf{5126}, 2008, 373--385.

\bibitem{20} G. van Drimmelen, \textit{Satisfiability in alternating-time temporal logic}, Proceedings of the 18th Annual IEEE Symposium of Logic in Computer Science, 2003, 208--217.

\bibitem{21} M. Kacprzak and W. Penczek, \textit{Unbounded model checking for Alternating-Time Temporal Logic}, Proceedings of the Third International Joint Conference on Autonomous Agents and Multiagents Systems, 2004, 646--653.

\bibitem{22} L. Gordeew, E. H. Haeusler, \textit{Proof Compression and NP Versus PSPACE}, Studia Logica, \textbf{107}:1, 2019, 55--83; 

\bibitem{23} L. Gordeew, E. H. Haeusler, \textit{Proof Compression and NP Versus PSPACE II}, Bulletin of the Section of Logic, \textbf{49}:3, 2020, 213--230.


\end{thebibliography}
\end{document}